\documentclass[
tightenlines,
preprintnumbers,
nofootinbib,
twocolumn,
 amsmath,amssymb]{revtex4-2}
\usepackage{amsmath}
\DeclareMathOperator*{\argmax}{arg\,max}

\usepackage{amssymb}
\usepackage{bbold}
\usepackage{bm}
\usepackage{color,graphicx}
\usepackage{slashed}
\usepackage{comment}
\usepackage{hyperref}
\usepackage{subcaption}
\usepackage{makecell}
\usepackage[dvipsnames]{xcolor}
\usepackage{pifont}
\usepackage{caption}
\usepackage{float}
\usepackage{array}
    \newcolumntype{P}[1]{>{\centering\arraybackslash}p{#1}}
    \newcolumntype{M}[1]{>{\centering\arraybackslash}m{#1}}

\DeclareMathSymbol{\mathbbE}{\mathord}{AMSb}{"45}
\newcommand{\ex}{\mathbbE}

\DeclareMathSymbol{\mathbbH}{\mathord}{AMSb}{"48}
\DeclareMathSymbol{\mathbbR}{\mathord}{AMSb}{"52}

\begin{document}

\renewcommand\labelitemi{$\vcenter{\hbox{\tiny$\bullet$}}$}

\preprint{ANL-192025}

\title{Anomalous electroweak physics unraveled via evidential deep learning}

\author{Brandon Kriesten}
\email{bkriesten@anl.gov}

\author{T.~J.~Hobbs}
\email{tim@anl.gov}

\affiliation{
    High Energy Physics Division, Argonne National Laboratory, Lemont, IL 60439
}

\date{\today}

\begin{abstract}
The ever-growing ecosystem of beyond standard model (BSM) calculations and parametrizations has motivated the development of systematic methods for making
quantitative cross-comparisons over the wide range of possible models, especially with controllable uncertainties.
In this setting, the language of uncertainty quantification (UQ) furnishes useful metrics for assessing statistical overlaps and discrepancies among BSM and related models. In this study, we leverage recent
machine learning (ML) developments in evidential deep learning (EDL) for UQ to separate data (aleatoric) and knowledge (epistemic) uncertainties in a model-discrimination setting.
We construct several potentially BSM-motivated scenarios for the anomalous electroweak interaction (AEWI) of neutrinos with nucleons in deep inelastic
scattering ($\nu$DIS). These scenarios are then quantitatively mapped, as a demonstration, alongside Monte Carlo replicas of the CT18 PDFs used to calculate
the $\Delta \chi^{2}$ statistic for a typical multi-GeV $\nu$DIS experiment, CDHSW.  Our framework effectively highlights areas of model agreement and
provides a classification of out-of-distribution (OOD) samples. By offering the opportunity to quantitatively understand
model overlaps, the approach presented in this work can help facilitate efficient BSM model exploration and exclusion for future New Physics searches. 
\end{abstract}

\maketitle

%
\section{Introduction}
\label{sec:intro}

Following the completion of the standard model (SM), there has been an efflorescence of beyond SM (BSM) theory developments~\cite{Fox:2022tzz} and parallel empirical tests at high-energy
facilities.
As the landscape of BSM theory and searches has steadily grown in complexity, theoretical methods for discriminating among models and quantifying their differences using
robust statistics have enjoyed renewed interest. Problematically, BSM frameworks often have numerous model- or parametrization-specific degrees of freedom, making it difficult to establish a tangible, statistical representation in which to distinguish
or compare among the range of possible scenarios on a common footing.
While effective field theory (EFT)-based methods like SMEFT~\cite{Brivio:2017vri} attempt to ameliorate this issue by offering an operator basis of fixed dimension to which UV-complete models might be mapped, practical analyses generally assume a truncation in the EFT expansion and fit only a subset of the available operators.

As one favorable approach, it can be advantageous to reduce the dimensionality of parameter spaces spanned by BSM models to tractably calculable projections~\cite{Hallin:2024gmt}
wherein commonalities and distinctions among these models might be statistically evaluated relative to available HEP data.
In this study, we present a novel realization of this approach, leveraging recent advances in evidential deep learning (EDL) to statistically separate BSM models in a generalizable example involving simple parametric variations in electroweak theory; a critical aspect of the approach we demonstrate is the possibility of constructing mappings to inform such model discriminations with a notion of uncertainty quantification (UQ) in the model separation.
Quantifying statistical commonalities among BSM scenarios along the lines shown in our calculation below presents an opportunity to enhance the understanding of New Physics signatures in an apples-to-apples setting.

We demonstrate our EDL-based model separations in the specific instance of neutrino deeply inelastic scattering ($\nu$DIS).
Neutrino interactions play a valuable role in the HEP landscape~\cite{Ackermann:2022rqc}, spanning a wide kinematical range from GeV-scale accelerator-based oscillation searches, to
TeV-scale collider studies, to extraterrestrial ultra high-energy signals from astrophysical neutrinos measured in the thousands of EeV. Meanwhile, charge-current (CC) $\nu$DIS not only informs contemporary knowledge of the structure of QCD matter within the SM, but is also sensitive to a variety of BSM scenarios~\cite{Arguelles:2022tki}, including generalized neutrino interactions (GNI), leptoquarks, and hidden dimensions. Furthermore, the experimental observation that neutrino flavor eigenstates mix, and thus neutrinos have non-zero mass, demonstrates the incompleteness of the SM in the neutrino sector. 
A traditional setting for exploring the interplay of QCD effects and possible BSM signatures has been the technique of global analysis. In the case of QCD analyses of parton
distribution functions (PDFs)~\cite{Alekhin:2017kpj,Hou:2019efy,Bailey:2020ooq,NNPDF:2021njg}, which have often sought to extract sensitive information on, {\it e.g.} the strange content of the proton from $\nu$DIS, calculations often involve
a range of subtleties~\cite{Accardi:2016qay,Hou:2019efy,Bailey:2020ooq,NNPDF:2021njg,Risse:2023wxr,Candido:2023utz}. For example, because of the weak nature of the neutrino-nucleon interaction, heavy targets such as lead, iron, or tungsten are used to achieve higher statistics; therefore, nuclear corrections are needed to separate the pure collinear proton PDF from nuclear effects. Furthermore, QCD global analyses involving neutrino data are complicated by possible tensions between $\nu$DIS data and charged-lepton DIS~\cite{Muzakka:2022wey}. These various challenges improve, but do not vanish, when correlated uncertainties are taken into account; this fact, together with the potential BSM sensitivity of $\nu$DIS leaves strong motivation for continuing refinement of the (B)SM theory for $\nu$DIS. Given these persistent questions, many new searches are planned such as the Forward Search Experiment at the LHC (FASER$\nu$) \cite{FASER:2018bac, FASER:2019dxq, FASER:2021mtu, Feng:2022inv} and the near detector of the deep underground neutrino detector, DUNE-ND, \cite{DUNE:2016hlj,DUNE:2022aul} to search for potential BSM signatures.   
To confront this complicated landscape, we leverage recent developments in evidential deep learning (EDL) and related UQ techniques (see Ref.~\cite{DBLP:journals/corr/abs-2110-03051} for an overview), not only to classify the BSM models in question, but also to define uncertainty metrics which we repurpose as a measure of model distinction and overlap. In particular, we use Dirichlet Prior Networks (DPNs) \cite{malinin2018predictiveuncertaintyestimationprior} for this classification (or model discrimination) task to simultaneously separate the BSM scenarios, measure the degree of statistical overlap among them, and quantify the lack of knowledge regarding these scenarios in an effective model space. Additionally, the DPN can determine when a sample is completely out-of-distribution (OOD) --- indicating that no other model samples resemble it --- through a separate statistic.
We connect these challenges to three types of uncertainties: aleatoric uncertainty, which arises from data label overlap; epistemic uncertainty, stemming from a lack of knowledge or data in particular regions; and distributional uncertainty, which indicates whether a sample lies outside the training distribution (and is therefore OOD) --- effectively defining the boundaries of in-distribution sampling. These uncertainty measures can be written in closed form through the parameters of the Dirichlet distributions, facilitating efficient computation of uncertainty metrics in high-dimensional domains.
We note that these methods have the potential to complement other ML approaches, such as Gaussian Mixture Models~\cite{Yan:2024yir}.

In this manuscript, we address the challenge of disentangling BSM signatures by using an effective latent space of low dimensionality --- specifically, a two-dimensional plane of $\Delta \chi^{2} / N_\mathrm{pt}$ ({\it i.e.}, per datum) for $Q^{2}\! \le\! 10$ and $Q^{2}\! >\! 10$ GeV$^{2}$, separating the theoretical description of $\nu$DIS into low- and high-$Q^2$ regions. The novelty of our approach lies in providing quantifiable measures of model separation, overlap, and identification of OOD behavior in this basis. Our approach successfully classifies BSM models with a high degree of accuracy and provides robust uncertainty estimates, demonstrating its effectiveness in distinguishing models of approximate parametric similarity.

In terms of the practical aspects of our calculation, we construct three distinct electroweak parameter combinations based on deviations of charge-current electroweak couplings, which we generically term {\it anomalous electroweak interaction}
(AEWI) scenarios, and deploy evidential learning methods to explore statistical relationships in distinguishing among them.
These three AEWI scenarios emerge from randomly varying the elements of the CKM matrix according to their measured uncertainties to simulate potential signatures of non-standard neutrino interactions in $\nu$DIS.
Ensembles of theory predictions may then be obtained by assuming PDF Monte Carlo replicas computed from the NNLO CT18 Hessian sets~\cite{Hou:2019efy,Hou:2016sho}, taken in conjunction with these AEWI scenarios.
We project the resulting theoretical predictions for
a typical $\nu$DIS data set, CDHSW~\cite{Berge:1989hr}, into the two-dimensional space noted above to train our DPN framework.
The techniques introduced here for re-purposing UQ methods in BSM-sensitive classification can be extended to a variety of phenomenological studies where model discrimination and UQ are essential. Such studies include the phenomenology of collinear PDF fits using generative machine learning (ML) models constrained by lattice-calculated Mellin moments~\cite{Kriesten:2023uoi}. Additionally, these uncertainty metrics are useful for fits of deeply virtual exclusive processes~\cite{Guo:2022cgq,Almaeen:2022imx,Almaeen:2024guo}
and the multi-dimensional quantum correlation functions like generalized parton distributions (GPDs)~\cite{Kriesten:2019jep,Kriesten:2021sqc} to which such data are sensitive.
This approach has the potential to enhance phenomenological benchmarking studies~\cite{PDF4LHCWorkingGroup:2022cjn,Jing:2023isu} of PDFs by integrating model-similarity metrics, derived from UQ, into latent-space analyses. By employing these congruency measures as weights, in addition to ML explainability techniques introduced in Ref.~\cite{Kriesten:2024are}, we can more accurately represent model differences through Euclidean distances. This allows a more nuanced evaluation of benchmarking performance and provides a potential metric for easing tensions between PDF phenomenological fits. A comprehensive study on these applications is well-motivated and reserved for future work.

The rest of the manuscript is organized as follows: in Sec. \ref{sec:bsm} we contextualize our AEWI assumptions within EFTs as used to parametrize BSM scenarios before
discussing their implementation and interplay with standard model aspects of $\nu$DIS phenomenology in Sec.~\ref{sec:pheno}.
Sec.~\ref{sec:edl-theory} provides an in-depth discussion on the technical aspects of EDL by establishing a basis of standard Bayesian methods
in Sec.~\ref{sec:bayesian-methods}, constructing the DPNs in Sec.~\ref{sec:dpn-theory}, and defining the uncertainty measures we use in our
analysis in Sec.~\ref{sec:uncertainty-metrics}. We give an overview of our results in the classification of these models in Sec.~\ref{sec:results} and conclude in
Sec.~\ref{sec:conclusions}, including a brief outlook regarding possible extensions of this work.
A short Appendix provides the explicit parameter values associated with our AEWI scenarios.

%

\section{Beyond Standard Model context}
\label{sec:bsm}

Contemporary BSM phenomenology at colliders is significantly motivated by the assumption that the New Physics is characterized by
very heavy masses, $\Lambda\! \gg\! M_{W,Z}$, beyond the electroweak scale.
The corresponding space of BSM models consistent with this hypothesis entails UV completions well beyond currently accessible kinematics at terrestrial experiments; this
landscape of BSM models includes many scenarios, such as minimal supersymmetric models, composite-Higgs models, and
frameworks introducing novel particles like leptoquarks, $Z'$ bosons, or extra-dimensional theories.
Across this range of BSM scenarios is the limitation that the resonant production of undiscovered states likely lies beyond the reach of contemporary colliders.

At the same time, virtual processes mediated by heavy (BSM) degrees-of-freedom may still imprint nonresonant signatures at lower
energies, signaled by subtle deviations~\cite{Carrazza:2019sec,Greljo:2021kvv,Gao:2022srd} of precision measurements from SM 
baselines computed with the latest theory accuracy.
As these deviations are likely to be small, a natural approach to constrain BSM is to formulate the possible effects with
as minimal model dependence as possible, often in the form of EFTs~\cite{Weinberg:1978kz,Buchmuller:1985jz,Leung:1984ni}.
The most widely adopted EFT for TeV-scale physics has been SMEFT~\cite{Brivio:2017vri}.
SMEFT exploits the large separation between the hypothetical BSM scale and the scales characterizing
lower-momentum processes at colliders to formulate non-standard interactions purely in terms of the known field
content of the SM. The resulting ansatz can then be organized systematically into a complete basis~\cite{Grzadkowski:2010es} of
operators of given dimension $d$, $\mathcal{O}^d_i$, with the presence of BSM quantified in a gauge-invariant
manner via the corresponding Wilson coefficients, $C_i$.
This paradigm corresponds to an expanded Lagrangian of which the first term is the purely SM contribution:
\begin{align}
\label{eq:smeft}
	\mathcal{L}\ &=\ \mathcal{L}^{(d=4)}_\mathrm{SM} \\ 
	&+\, \sum_i {C^{(6)}_i \over \Lambda^2} \mathcal{O}^{(d=6)}_i\,
                                      +\, \sum_i {C^{(8)}_i \over \Lambda^4} \mathcal{O}^{(d=8)}_i\, +\, \cdots\ . \nonumber
\end{align}
For a given choice of operator dimension, SMEFT fully parametrizes non-standard interactions,
allowing constraints to BSM [signaled by $C_i\! \neq\! 0$ in Eq.~(\ref{eq:smeft})] through fits~\cite{Berthier:2015gja,Ellis:2018gqa,Ethier:2021ydt,Bissolotti:2023vdw,Giani:2023gfq} of high-energy data
which are agnostic relative to any particular UV-complete model.
Crucially, specific UV models can be mapped onto a given SMEFT operator basis such that patterns among fitted
Wilson coefficients might suggest~\cite{Henning:2014wua} an underlying BSM interpretation.
To describe BSM effects in an EFT formulation well below the EW scale it is possible to define a low-energy theory in which the electroweak has been
integrated away in addition to the $\Lambda$-scale interactions parametrized by SMEFT in Eq.~(\ref{eq:smeft}).
Within such a weak EFT (WEFT)~\cite{Buchmuller:1985jz,Grzadkowski:2010es,Jenkins:2017jig}, one can then compute systematic matchings at the weak
scale, $\mu\! \approx\! M_W$, with respect
to the SMEFT operators discussed above and evaluate corresponding runnings of the Wilson coefficients~\cite{Jenkins:2013zja,Jenkins:2013wua,Alonso:2013hga},
ultimately constraining these to experimental results at a given scale~\cite{Falkowski:2019xoe}.
In this fashion, a WEFT Lagrangian may parametrize generalized neutrino interactions (GNI),
including left- and right-handed as well as scalar, pseudo-scalar, and tensor couplings~\cite{Falkowski:2021bkq}; schematically, of
greatest relevance to the neutrino-quark interactions in charge-current DIS are the left-handed interactions, for which
a subset of the WEFT Lagrangian may be written
\begin{align}
\label{eq:WEFT}
&\mathcal{L}_{\text{WEFT}} \supset  \\
	&-\frac{2 V_{ij}}{v^2}
\Bigg\{ \left[ 1 + \epsilon_L^{ij} \right]_{ab} (\bar{u}^i \gamma^\mu P_L d^j)(\bar{\ell}_a \gamma_\mu P_L \nu_b) + \text{h.c.} \Bigg\}\ , \nonumber
\end{align}
where $\epsilon^{ij}_{L} = \epsilon_{ij}$ represent Wilson coefficients and we drop the left-handed notation for the remainder of this study. In
addition, $1/v^2 = \sqrt{2}G_F$ in Eq.~(\ref{eq:WEFT}), $(ab)$ and $(ij)$ represent lepton- and quark-generation indices, respectively, and the projector
is $P_L = [1-\gamma_5]/2$.
We note that additional Lorentz structures can contribute to the WEFT Lagrangian, but we neglect these for the present analysis, which concentrates
singly on charge-current DIS.
For this study, we therefore reiterate that it is possible to descend a chain of successive matchings, from specific model-dependent scenarios
with UV completions at $\Lambda \gg M_W$, to projections onto combinations of SMEFT operators, $\mathcal{O}_i$, which may themselves be matched to the
WEFT operators in Eq.~(\ref{eq:WEFT}) and run according to the associated Wilson coefficients, $\epsilon_{ij}$. We point out that the scenario
$\epsilon_{ij} = 0$ simply recovers the SM neutrino-quark interactions. In this work, we do not presuppose any specific BSM model or combination
of EFT operators, but rather notice the structure implicit in Eq.~(\ref{eq:WEFT}): namely, for the left-handed neutrino-quark operators which
contribute to SM $\nu$DIS, WEFT effectively renormalizes the interaction and we take
\begin{equation}
	{V_{ij} \over v^2} \to {V_{ij} \over v^2} \left[ 1 + \epsilon_{ij} \right] \delta_{ab}\ ,
	\label{eq:shift1}
\end{equation}
where we neglect any off-diagonal lepton-flavor effects ($\sim\delta_{ab}$). 
In this calculation, we restrict ourselves to $\nu$DIS as noted above, and therefore neglect the array of additional processes which might probe
the full basis of WEFT operators partly represented in Eq.~(\ref{eq:WEFT}). On this logic, we take the further step of absorbing any parton-level
signatures of non-standard physics in Eq.~(\ref{eq:shift1}) into shifted CKM matrix elements
\begin{equation}
	V_{ij} \to V^\prime_{ij} = \Delta_{ij}\, \cdot\, V_{ij}\ ,
	\label{eq:shift2}
\end{equation}
in which $\Delta_{ij}$ represent effective reweightings of the flavor-dependent neutrino-quark interaction into which we enfold 
possible effects of non-standard neutrino-parton interactions. We note that, while the charge-current interaction shifts of
Eq.~(\ref{eq:shift2}) can be justified within the SMEFT $\to$ WEFT BSM parametrizations discussed above, the practical
implementation below does not directly depend on the operator structure of these EFTs beyond the minimal assumptions noted
above regarding the left-handedness of the non-standard interaction. For this reason, we refer to the parametrizations obtained below
more generally as {\it anomalous electroweak interaction} (AEWI) scenarios in that these are obtained as deviations in charge-current
couplings without regard to an assumed BSM parametrization. We note, however, that top-down BSM models or SMEFT operators could ultimately
be curated and identified with shifts such as those deployed below.

\begin{figure*}[th]
    \centering
    \includegraphics[width=0.48\linewidth]{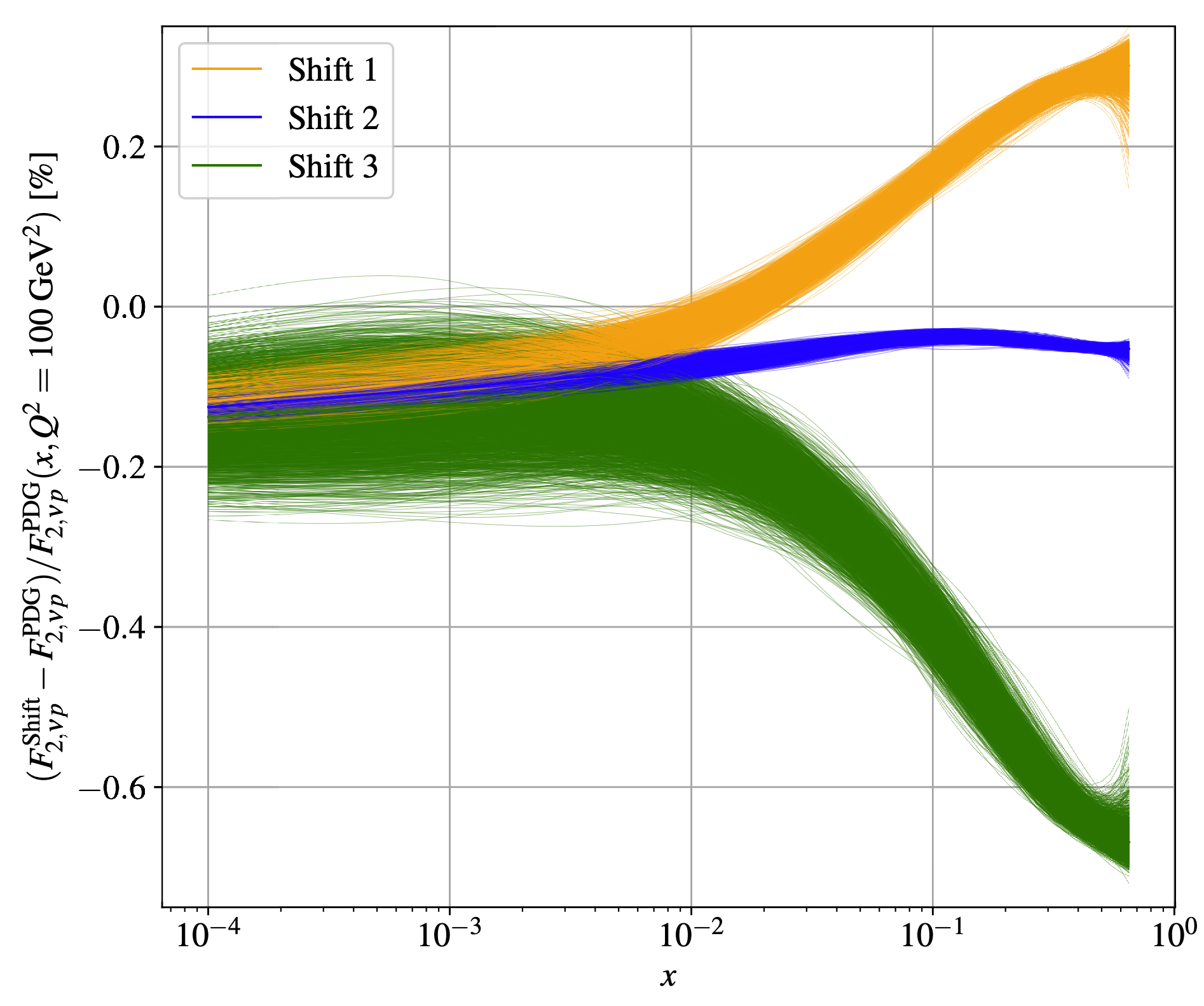}\ \ \
    \includegraphics[width=0.48\linewidth]{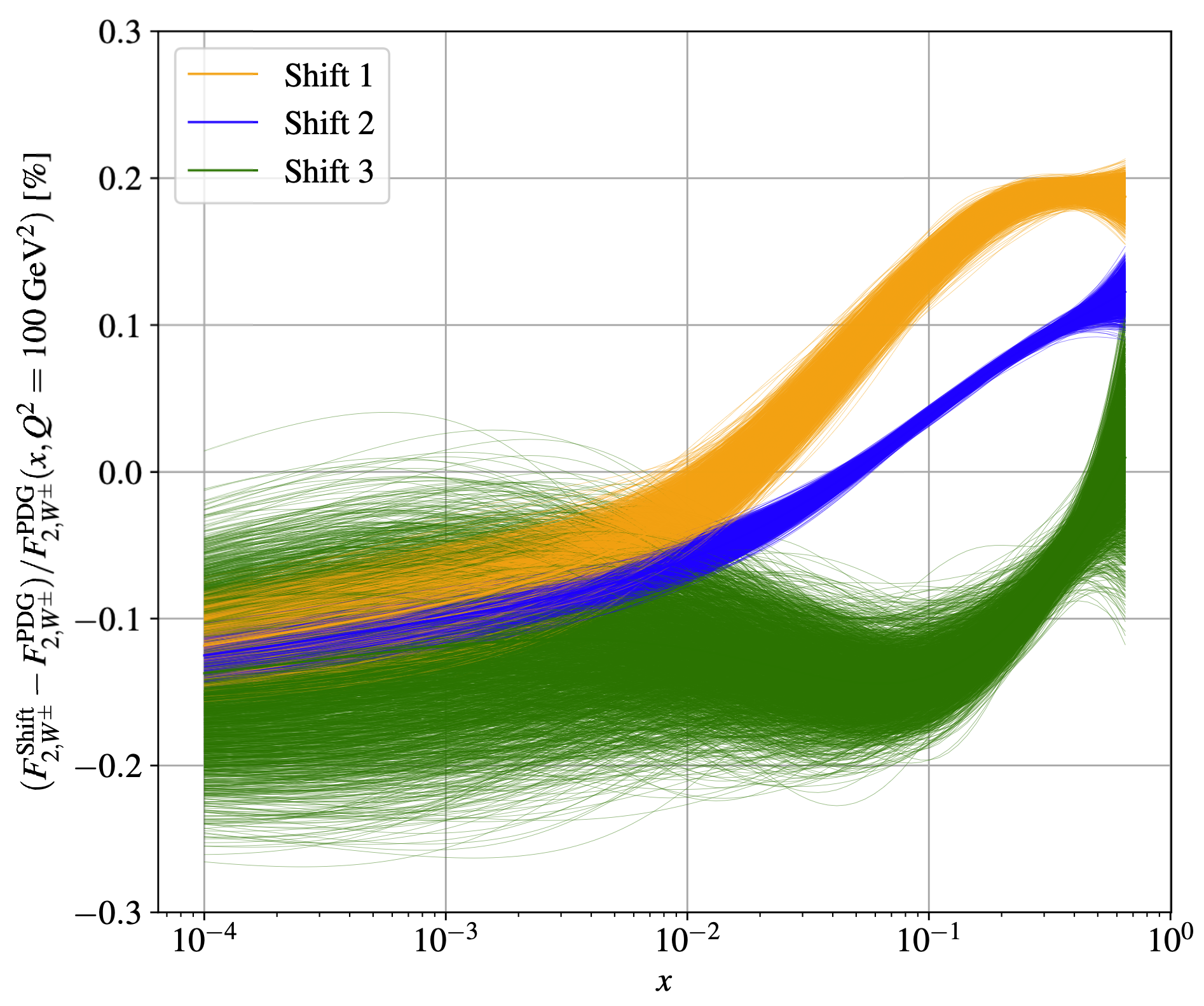}
	\caption{(Left) The fractional shifts in the neutrino-proton structure function, $F^{W^+}_2(x,Q^2)$ as induced
	by three randomized shifts in the CKM matrix elements corresponding to the AEWI scenarios discussed in Sec.~\ref{sec:bsm}.
	(Right) The corresponding fractional shifts on the charge-averaged (anti-)neutrino combination, $F^{W^\pm}_2(x,Q^2)$, as extracted
	by, {\it e.g.}, CDHSW. We plot these deviations at a fixed scale, $Q^2\! =\! 100\,\mathrm{GeV}^2$.}
    \label{fig:dis_sf}
\end{figure*}

In the end, the presence of the AEWI shifts leads to an effective CKM matrix, $V^\prime_{ij} = \Delta_{ij} V_{ij}$, 
\begin{eqnarray}
    V^\prime_{ij} &=& \begin{bmatrix}
        \Delta_{ud} \cdot V_{ud} & \Delta_{us} \cdot V_{us} & \Delta_{ub} \cdot V_{ub} \\
        \Delta_{cd} \cdot V_{cd} & \Delta_{cs} \cdot V_{cs} & \Delta_{cb} \cdot V_{cb} \\
        \Delta_{td} \cdot V_{td} & \Delta_{ts} \cdot V_{ts} & \Delta_{tb} \cdot V_{tb}   
\end{bmatrix}\ .
\label{eq:Vprime}
\end{eqnarray}
In Eq.~(\ref{eq:Vprime}) above, the various flavor-dependent shifts are $\sim\! 1$ due to the relative smallness of the $\epsilon_{ij}$
coefficients.
This parametrizes the New Physics as subtle changes in the EW mixing between different generations of quarks in the charged-current weak interactions,
within uncertainties.
As discussed in more detail in Sec.~\ref{sec:pheno} below, having formulated $V^\prime_{ij}$, we can then implement randomized shifts in the CKM-weighted
neutrino-quark interactions by choosing $\Delta_{ij}$ to deviate from current PDG-preferred values~\cite{ParticleDataGroup:2024cfk} by approximate ``discovery-level'' margins of $\sim$several
sigma.
We enforce approximate preservation of CKM unitarity within uncertainties~\cite{Descotes-Genon:2018foz}.

We implement these shifts to electroweak physics in the context of $\nu$DIS, which has the added complication of subtle
cross-dependence between electroweak interactions and QCD in the form of quark-gluon structure of nucleons and nuclei. We
discuss the associated implementation and issues in Sec.~\ref{sec:pheno} below, highlighting the challenge of separating
BSM effects from variations in the parton distribution functions (PDFs) of the proton and nuclei and consequences for theoretically
describing a typical $\nu$DIS data set (CDHSW). Having established these aspects of the phenomenology, we then turn
to the use of EDL in discriminating among several AEWI scenarios in Sec.~\ref{sec:edl-theory}.

%

\section{Phenomenology of $\nu$DIS}
\label{sec:pheno}

BSM-induced shifts to charge-current interactions as discussed above have the potential to influence a range of physical observables. In this study, given the intriguing question of interference between BSM effects and standard model theory calculations, we select $\nu$DIS
and concentrate our EDL-based model discrimination calculations on AEWI-driven variations in $\nu$DIS structure functions.
$\nu$DIS has attracted sustained interest due to its potential sensitivity to BSM signatures, as discussed above, through its dependence on fundamental parameters of the electroweak sector.
A characteristic example of a precision extraction of electroweak parameters from $\nu$DIS is the anomalous determination of $\sin^2\theta_W$ by the NuTeV Collaboration~\cite{NuTeV:2001whx}, which suggested a possible $\sim 3\sigma$ deviation from the standard model.
The theoretical interpretation of the reported NuTeV anomaly has received considerable attention (see, {\it e.g.}, Refs.~\cite{Pumplin:2002vw,Miller:2002xh,Bentz:2009yy}) as it depends on precise control over an array of (non)perturbative QCD and nuclear effects --- a fact which reflects the role of QCD uncertainties.
This reality derives from the fact that $\nu$DIS probes unique flavor currents in QCD matter in addition to its sensitivity to short-distance aspects of the electroweak interaction.
Both elements are reflected in perturbative QCD and electroweak calculations of $\nu$DIS, as well as
in global analyses of the type noted in Sec.~\ref{sec:intro}, which implement this theory.

The reduced cross section for neutrino (or antineutrino) scattering on nucleons as measured in charge-current DIS depends on corresponding structure functions, $F^{W^{+/-}}_{2,3,L}$, for either $W^+$ or $W^-$ exchange, respectively, which in turn depend on both electroweak couplings and PDFs. Namely,
\begin{align}
    \frac{d^2 \sigma^{W^{+/-}}}{dx \, dy} = &{1 \over 8 \pi v^4} {Q^2 \over xy} \left( {M^2_W \over Q^2 + M^2_W}\right)^2 \\
    &\times \left[ Y_+ F_2^{W^{+/-}} \pm Y_- x F_3^{W^{+/-}} - y^2 F_L^{W^{+/-}} \right]\ , \nonumber
\end{align}
where $x, y$, and $Q^2$ are the usual DIS invariants, $Y_\pm = 1 \pm (1-y)^2$, and we assume the virtuality of the process is large relative to the nucleon mass, $Q^2\! \gg\! M^2$. The dependence on PDFs enters the structure functions via convolution with coefficient functions, $C_{i,j}$~\cite{Blumlein:2014fqa}, representing the hard partonic scattering, where the charge-current structure functions may be factorized~\cite{Gao:2021fle} to all orders in $\alpha_s$ as
\begin{align}
\label{eq:fact}
    F(x, Q^2) &= \sum_m \sum_n \{ C_{m,n} \otimes \Phi_n \}(x, Q^2) \\
    C_{m,n}(z) &= C_{m,n}^{(0)} + a_s C_{m,n}^{(1)} + a_s^2 C_{m,n}^{(2)} + \mathcal{O}(a_s^3)\ , \nonumber
\end{align}
with $a_s = \alpha_s/4\pi$, and where the quantities $\Phi_n$ contain PDFs of flavor $n$. As the charge-current interaction is quark flavor-changing, Eq.~(\ref{eq:fact}) requires flavor sums over the initial- ($n$)  and final-state quark flavors ($m$), and dependence on the CKM quark mixings is therefore implicit.
In the LO 
expression of Eqs.~(\ref{eq:CCSF_1})--(\ref{eq:CCSF_2}), these perturbatively calculable Wilson coefficients are delta functions, $C^{(0)}_{m,n} = \delta(1-z)$, but become more
complicated at the NNLO accuracy used in the results below.
For interactions
with individual nucleons, the leading-order (LO) structure function for neutrino-proton scattering can be written as
\begin{align}
\label{eq:CCSF_1}
&F^{W^+}_2(x,Q^2) = \\
&2x \bigg\{ \sum_{j=d,s,b} |V^\prime_{uj}|^2 \bar{u}(x,Q^2) + \sum_{i=u,c,t} |V^\prime_{id}|^2 \,d(x,Q^2) + \cdots \bigg\}\ , \nonumber
\end{align}
while the corresponding expression for antineutrino scattering is
\begin{align}
\label{eq:CCSF_2}
&F^{W^-}_2(x,Q^2) = \\
&2x \bigg\{ \sum_{i=u,c,t} |V^\prime_{id}|^2 \bar{d}(x,Q^2) + \sum_{j=d,s,b} |V^\prime_{uj}|^2 \,u(x,Q^2) + \cdots \bigg\}\ ; \nonumber
\end{align}
above, the ``$\cdots$'' involve contributions from heavier $u$- and $d$-type quark generations and follow the same pattern of dependence on the CKM matrix elements.
For the sake of this analysis, the essential aspect revealed in Eqs.~(\ref{eq:CCSF_1})--(\ref{eq:CCSF_2}) is the combined dependence on
the AEWI-shifted parton-level electroweak couplings ($V^\prime_{ij}$) discussed in Sec.~\ref{sec:bsm} and the PDFs [$q(x,Q^2)$], which we have made explicit by
keeping the foregoing expressions at LO in QCD.
We implement the basic formalism above in computing theoretical predictions for neutrino-DIS measurements at next-to-next-leading order (NNLO) theory accuracy in $\alpha_s$ as noted.
In particular, in Sec.~\ref{sec:results}, we demonstrate the EDL methods of Sec.~\ref{sec:edl-theory} against the CDHSW $\nu$DIS data set, which extracted the effectively charge-neutral structure function combination,
\begin{equation}
\label{eq:cdhsw}
    F_{2}^{W^\pm}(x,Q^{2}) \equiv {1 \over 2}\, \bigg( F_{2}^{W^+}(x,Q^{2}) + F_{2}^{W^-}(x,Q^{2}) \bigg)\ ,
\end{equation}
from neutrino and antineutrino interactions on an Fe target. For the practical illustration of this study, we concentrate on the statistical discrimination of AEWI-induced variations in the presence of a fixed PDF and its associated uncertainty. We note, however, that this exercise can be generalized to additional variations in the assumed theory or related inputs, including simultaneous floating of the electroweak parameters and PDFs.
We compute theoretical predictions for the CDHSW data at NNLO using the \texttt{yadism} package~\cite{Candido:2024rkr} interfaced to CT18 PDFs (also fitted at NNLO), for which we generate 3000-member Monte Carlo replica ensembles using the \texttt{mcgen} code~\cite{Hou:2016sho}.
With these inputs, we compute families of predictions for the structure function combination of Eq.~(\ref{eq:cdhsw}), generating 3000 PDF-based replica predictions for a given choice of CKM mixings --- either the central PDG preferred values~\cite{ParticleDataGroup:2024cfk}, or one of three AEWI-shifted parameter sets, $V^\prime_{ij}$.
For the AEWI-induced shifts described qualitatively in Sec.~\ref{sec:bsm}, we take three concrete scenarios, each of which alters neutrino-nucleon cross-sections or structure functions in line with the formalism described above.
In particular, we take randomized shifts, but curate the magnitude of these shifts so as to produce $\lesssim\,$percent-level effects in the DIS structure functions.
We summarize the specific numerical values assumed for $V^\prime_{ij}$ in the Appendix.
In Fig.~\ref{fig:dis_sf}, we explicitly show the impact of the AEWI-shifted CKM parameters on the charge-current structure functions, plotting the shifted structure functions, $F^\mathrm{shift}_2 - F^\mathrm{PDG}_2$, normalized to the calculation based on nominal PDG values for $Q^2\! =\! 100\,\mathrm{GeV}^2$. In the left panel, we exhibit the shifts in the $F^{W^+}_2$ structure function of the proton, associated with neutrino-proton scattering; in the right panel, we show the analogous quantity for the CDHSW-relevant structure function combination, $F^\pm_2$ of Eq.~(\ref{eq:cdhsw}). In both cases, the AEWI shifts encoded in $V^\prime_{ij}$ produce a $\lesssim\! 1\%$ span in the high-$x$ shape of the structure function, particularly in the neutrino-nucleon case; the CDHSW combination somewhat mitigates this spread, which nonetheless remains clearly separated relative to PDF uncertainties.
Our EDL calculation ultimately entails evaluating $F_{2}^{W^\pm}$ of Eq.~(\ref{eq:cdhsw}) as plotted in Fig.~\ref{fig:dis_sf} (right) for the CDHSW data, and computing the corresponding $\chi^2$ for the full set of experimental points; we determine this overall $\chi^2$ for a given choice of the CKM elements and, on this basis,
evaluate
\begin{equation}
\label{eq:del-chi}
   {1 \over N_\mathrm{pt}}\, \Delta \chi^2 = {1 \over N_\mathrm{pt}} \big( \chi^2_\mathrm{PDG} - \chi^2_\mathrm{shift} \big) 
\end{equation}
relative to PDF uncertainties by assessing the $\Delta \chi^2$ of Eq.~(\ref{eq:del-chi}) above per MC replica; $\chi^2_\mathrm{PDG}$ assumes the nominal PDG quark mixings, and $\chi^2_\mathrm{shift}$ represents one of the three AEWI shift scenarios enumerated in the Appendix.
For the sake of applying the EDL model discrimination prescription discussed in Sec.~\ref{sec:edl-theory}, it is favorable to represent configurations of $\Delta \chi^2$ in a two-dimensional space; we construct these by segmenting the $\Delta \chi^2$ evaluations of Eq.~(\ref{eq:del-chi}) into contributions coming from low ($\le\! 10\, \mathrm{GeV}^2$) {\it vs}.~high ($>\! 10\, \mathrm{GeV}^2$) values of $Q^2$.
In addition to the practical benefit of representing the CDHSW data in a two-dimensional space for model classification, there are also phenomenological motivations for the segmentation we use; it has been observed~\cite{Hou:2019efy} that the CDHSW data exhibit an anomalous $Q^2$ scaling potentially at odds with DGLAP evolution --- an aspect reflected in the $Q^2$ dependence at lower momentum transfers. By explicitly separating the data into distinct $Q^2$ subsets, we quantify any possible systematic differences in the descriptions within each subset while examining how AEWI-driven shifts might influence these.
We also point out that fixed-target $\nu$DIS from Fe implies that the CDHSW data require knowledge of nuclear-medium effects in order to correct them to the nucleon-level interactions corresponding to the expressions above. For this demonstration, we concentrate on AEWI shifts assuming a base SM theory, and we therefore use the default nuclear settings in \texttt{yadism}, which simply weight free-nucleon structure functions by the appropriate $(A,Z)$ numbers. 
For more comprehensive studies of BSM models in a global analysis with the EDL methods discussed below, it would be valuable to implement and simultaneously vary a more representative range of nuclear-correction models.

In the end, the core calculation above illustrates the subtle but commonplace interplay between possible signatures of BSM physics as might be imprinted on electroweak parameters and the PDFs which are often fitted to $\nu$DIS or similar hadronic experiments.
In Fig.~\ref{fig:dis_sf}, we illustrate the fact that the AEWI-driven shifts in the neutrino-nucleon interaction effectively reshuffle the contributions to the structure function arising from flavor-dependent PDFs, a point which can be seen at LO from the expressions in Eqs.~(\ref{eq:CCSF_1})--(\ref{eq:CCSF_2}).
This suggests the potential for significant correlation between BSM-sensitive electroweak parameters and the PDFs themselves, which might conceivably absorb signatures of the AEWI effects when actively fitted.
This possibility has a parallel realization in analyses of BSM-sensitive EFTs from which our AEWI-scenarios might flow.
The SM-only ($C_i\! =\! 0$) assumptions characterizing typical PDF fits have mirror analogues in (SM)EFT analyses:
Wilson coefficients, $C_i$, extracted in SMEFT generally assume SM baselines with frozen PDFs, biasing
BSM sensitivity.
Thus, while we concentrate on model discrimination in a fixed-PDF realization in the current study, the EDL techniques to statistically separate different AEWI scenarios might be deployed more broadly in calculations which simultaneously vary the PDFs as well~\cite{Carrazza:2019sec,Gao:2022srd,Hammou:2023heg,Costantini:2024xae,Shen:2024sci,Wang:2024gvt} --- a fact which further motivates the results obtained below.

%

\section{Evidential Deep Learning for classification}
\label{sec:edl-theory}

In this section, we give an overview of the fundamentals of UQ for classification tasks through the lens of EDL. We introduce classification through a probabilistic ML perspective before describing the Bayesian deep learning framework underpinning much of UQ in the literature, and finally introduce the theory of EDL for the prior networks which we ultimately use in our analysis. We develop these aspects of our study keeping in mind that the task of classification discussed below is identifiable with statistical model discrimination; it is therefore closely relevant for distinguishing the predictions of specific BSM-driven AEWI scenarios of Sec.~\ref{sec:bsm}--\ref{sec:pheno} with respect to empirical data. We note that, for those wishing to bypass the mathematical discussion of the EDL formalism, the summary of uncertainty metrics given in Sec.~\ref{sec:uncertainty-metrics} can be consulted in-brief before moving to the Results and Conclusion of Sec.~\ref{sec:results}-Sec.~\ref{sec:conclusions}.

We start by defining a data set, $\mathcal{D} = \{\mathbf{x}^{(i)}, y^{(i)} \}_{i=1}^{N}$, of $N$ total data points generated from an unknown joint probability distribution, $p_{true}(\mathbf{x}, y)$, where $\mathbf{x}, y$ are random variables of the input and label spaces defined as $\mathbf{x} \in \mathbbR^{\mathcal{D}}$ and $y \in \{1, \dots, C \}$, respectively. The variable $y$ is often conveniently represented as a one-hot-encoded vector which takes the form $y \in \{\mathbf{0}, 1 \}^{C}$, that is, a vector of null entries with a single $1$ representing the positive class identification. A Bayesian classification model learns a functional probabilistic mapping, $f_{\theta}: \mathbbR^{\mathcal{D}} \rightarrow \{\mathbf{0}, 1 \}^{C}$, in which the learned probability distribution is defined as $\hat{\pi} = \sigma(f(\mathbf{x},\theta))$, such that:
\begin{eqnarray}
    p(y | \mathbf{x}, \theta) = \text{Cat}\left( y  | \sigma[ f(\mathbf{x}, \theta) ] \right) \ ,
\end{eqnarray}
where $\sigma$ is the standard Softmax function. The ML model learns to approximate a categorical multinomial distribution,
\begin{eqnarray}
    \text{Cat}(y|\hat{\pi}) &=& p(y = c | \mathbf{x}^{(i)}, \theta) \nonumber \\
    &=& \prod_{c=1}^{C} \pi_{c}^{\mathbb{1}(y^{(i)}=c)}\ ,
\end{eqnarray}
by predicting a vector of probabilities per class with the predicted label given as 
\begin{eqnarray}
    \hat{y} = \argmax_{c} \hat{\pi} \ .
\end{eqnarray}
This forms the basis of probabilistic ML wherein the neural network algorithm learns the underlying data distribution by approximating parameters of a given probabilistic distribution --- in the case of classification, a categorial distribution.

\subsection{Standard Bayesian Methods}
\label{sec:bayesian-methods}
A measurement of uncertainty in ML-based model predictions comes from the predictive posterior distribution, which is constructed both from the conditional likelihood, $p(y | \mathbf{x}^{*}, \theta)$, as well as the  Bayesian posterior over model parameters, $p(\theta | \mathcal{D})$. This is a form of Bayesian model averaging where the predicted label is weighted by how likely the model is:
\begin{eqnarray}
    p(y | \mathbf{x}^{*},\mathcal{D}) &=& \int p(y | \mathbf{x}^{*},\theta)\, p(\theta | \mathcal{D}) d\theta\ ,
\end{eqnarray}
where $x^{*}$ denotes a representative sampling of the input data set which is not seen during training.
Training a classification model via maximum likelihood estimation is equivalent to minimizing the negative log-likelihood (NLL); namely, the optimal parameters of the classification model are determined by minimizing the expectation value of the NLL over the true unknown data distribution, $\ex_{p_{true}(\mathbf{x},y)}[\cdot] \approx \frac{1}{N}\sum_{i}[\cdot]$, represented as
\begin{eqnarray}
    \theta^{*} 
    &=& \underset{\theta}{\text{argmin}}\Bigg( - \ex_{p_\mathit{true}(\mathbf{x},y)}\sum_{c=1}^{C} {\mathbb{1}(y^{(i)} = c)} \nonumber \\
    && \hspace{2.5cm} \times \ln{p(\hat{y} = c| \mathbf{x}^{(i)}, \theta)} \Bigg)\ .
\end{eqnarray}
This training procedure not only ensures that the conditional predictive distribution approximates the empirical distribution of the data, but it also  approximately factorizes the predictive posterior into two pieces. This can be seen by taking the expression for the expectation value over the NLL and rewriting it using the definitions of the KL divergence, $D_{KL}(\cdot)$, and the entropy, $\mathbbH(\cdot)$ \footnote{The derivation of this separation of the predictive posterior relies on the assumption of an infinite data set, $\mathcal{D}$, such that $p_\mathit{true}(\mathbf{x},y)$ is contained in the training set. This is why we state `approximately' separates.},
\begin{widetext}
\begin{eqnarray}
    \ex_{p_{true}(\mathbf{x},y)}\left[\mathcal{L}^\mathrm{NLL}(y,\mathbf{x},\theta) \right] &=& \ex_{p_{true}(\mathbf{x})}  \Bigg[ D_{KL}\Big( p_{true}(y| \mathbf{x}) \Big \| p( y| \mathbf{x}, \theta)\Big) - \mathbbH \Big(p_\mathit{true}(y| \mathbf{x})\Big) \Bigg]\ ,
\end{eqnarray}
\end{widetext}
where the first term, $D_{KL}\Big( p_\mathit{true}(y| \mathbf{x})) \Big \| p( y| \mathbf{x}, \theta)\Big)$, represents the {\it epistemic} or knowledge uncertainty, and the second term, $\mathbbH \Big(p_\mathit{true}(y| \mathbf{x})\Big)$, is the {\it aleatoric} or data uncertainty. Notice that the epistemic uncertainty is reducible through improved training procedures and more representative training data due to its dependence on the model parameters, $\theta$, but the aleatoric uncertainty is set by the true underlying and unknown probability distribution from which the data are drawn and therefore cannot be reduced with the further inclusion of training data.

In Bayesian methods, one must sample the model posterior many times and make predictions using an ensemble of models, $\mathcal{M}$, where we can write the ensemble posterior as $\{p(y | \mathbf{x}^{*}, \mathcal{M}^{(m)} ) \}_{m=1}^{M}$.
The predictive posterior can then be written in terms of this ensemble of models,
\begin{eqnarray}
p(y | \mathbf{x}^{*}, \mathcal{D}) = \ex_{p(\mathcal{M} | \mathcal{D})}\big[p(y | \mathbf{x}^{*}, \mathcal{M}) \big]\ . 
\end{eqnarray}
The ensemble posterior, $p(\mathcal{M} | \mathcal{D})$, is computationally difficult to evaluate for neural networks. Although there are some examples of methods that approximate the posterior through variational inference, such as Bayes by Backprop~\cite{blundell2015weightuncertaintyneuralnetworks}, in the end, the ensemble of models must be trained and used to predict on --- a computationally inefficient task.

\begin{figure*}
    \centering
    \includegraphics[width=2\columnwidth]{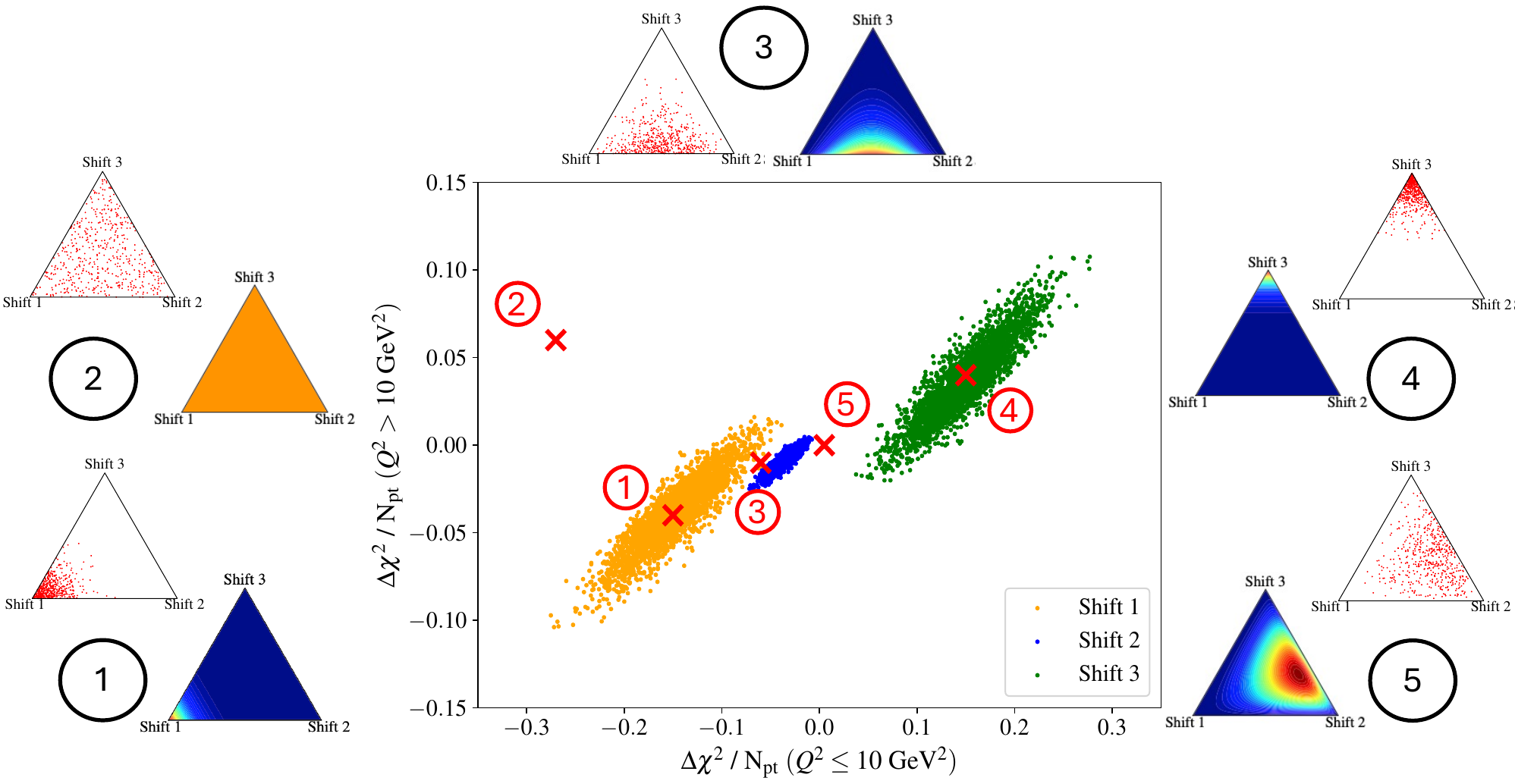}
    \caption{As a demonstration, we provide five representative samplings of Dirichlet Prior Network predictions based on training to three AEWI-shifted scenarios for the electroweak couplings, $V^\prime_{ij}$, and the CT18 PDF set; predictions are dimensionally reduced to a calculated $\Delta \chi^{2} / N_\mathrm{pt}$  statistic for $Q^{2}\! >\! 10$ GeV$^{2}$ and $Q^{2}\! \le\! 10$ GeV$^{2}$ on the CDHSW $\nu$DIS data set. For each of the five selections within the $\Delta \chi^2$ plane, the left-hand simplexes (red points) are 500 samples of the Dirichlet distribution while the right-hand simplexes show the full, contoured distribution.}
    \label{fig:dpn_demo}
\end{figure*}

%
\subsection{Dirichlet Prior Networks}
\label{sec:dpn-theory}

DPNs were introduced to solve the challenges discussed above by exploiting the fact that, in a single forward pass, a DPN approximates the previously intractable Bayesian ensemble posterior by predicting the parameters of the conjugate prior which produces the ensemble. This can be written schematically by factorizing the predictive posterior into an additional marginalization over the distribution, $\mu$, representing the choice of prior probability distribution:
\begin{eqnarray}
    p(y | \mathbf{x}^{*},\mathcal{D}) &=& \int \int p(y|\mu)\, p(\mu | \mathbf{x}^{*},\theta)\, p(\theta | \mathcal{D}) d\theta d\mu\ .\ \
\end{eqnarray}
We can assume a point estimate for the predicted model parameters in which $p(\theta | \mathcal{D}) = \delta(\theta - \hat{\theta})$, where $\hat{\theta}$ is the estimated value of the ML model's parameters based on observed data and $\theta$ is the unknown truth value of the model parameters.
We can then marginalize over the model parameters yielding
\begin{eqnarray}
    p(y | \mathbf{x}^{*},\mathcal{D}) &=& \int  p(y|\mu)\, p(\mu | \mathbf{x}^{*},\hat{\theta}) d\mu \ ,
\end{eqnarray}
where now $p(\mu | \mathbf{x}^{*}, \mathcal{D}) \approx p(\mu | \mathbf{x}^{*}, \hat{\theta}) = p(\mu ; \alpha = f(\mathbf{x}^{*},\hat{\theta}))$ is the predictive posterior distribution over the assumed prior. In the case of classification, the conjugate prior to the categorical multinomial distribution is the multinomial Dirichlet defined by concentration parameters, $\alpha$:

\begin{eqnarray}
    p(\mu ; \alpha) =  \frac{\Gamma(\alpha_{0})}{ \prod_{c=1}^{C}\Gamma(\alpha_{c})}\prod_{c=1}^{C}\pi_{c}^{\alpha_{c}-1} \ ,
\end{eqnarray}
 where $\alpha_{0} = \sum_{c=1}^{C}\alpha_{c}$, $\Gamma(\cdot)$ is the Gamma function, $C$ is the total number of classes, and $\pi_{c}$ remains a vector of class probabilities.  If we exponentiate the output scores of the neural network, {\it i.e.}, $S_{c} = f_{c}(x^{*},\hat{\theta})$, we then have $\alpha_{c} = e^{f_{c}(x^{*},\hat{\theta})}$, and the expectation of the predictive posterior for a particular example is related to the softmax function,
\begin{eqnarray}
    \frac{\alpha_{c}}{\alpha_{0}} &=& \left[ \frac{e^{S_{c=1}}}{\sum_{c=1}^{C} e^{S_{c}}}, \dots , \frac{e^{S_{c=C}}}{\sum_{c=1}^{C} e^{S_{c}}}\right]\ .
\end{eqnarray}
By learning the Dirichlet prior, we therefore can construct a distribution of likelihood functions representing classification scores. This not only allows us to make classification predictions (based on where the highest density of predictions lies in the $D$-simplex)
but also gives us a measure of uncertainty due to the spread and location of the density of predictions. Discrete probabilities over $C$ number of classes can be represented on a simplex of dimension $D\! =\! C-1$, meaning that the distributions obey the following properties: $\sum_{c=1}^{C} \pi_{c} = 1; \pi_{c} \in [0,1]$ for all $c \in \{1, \dots, C \}$. For example, a categorical distribution representing probabilities for classification of three classes can be represented on a two-dimensional simplex with three points, or a triangle. The categorical distribution can be plotted on this simplex using a barycentric coordinate transformation.

%
\subsection{Uncertainty Metrics}
\label{sec:uncertainty-metrics}
An advantage of the EDL statistical theory developed in the previous subsection is the fact that it can be deployed to further investigate UQ aspects of the AEWI model discrimination.
In particular, we can understand the origins of the classification model uncertainty by decomposing the total error into its distinct aleatoric and epistemic contributions via closed-form expressions in terms of the Dirichlet concentration parameters. These uncertainty metrics emerge from an information theoretic foundation in which entropy encapsulates the total uncertainty of a sampling. A low-entropy sampling indicates that the distribution is peaked along a specific class while a high-entropy instance is associated with a flat, uniform distribution. To express these metrics in closed-form expressions in terms of the Dirichlet concentration parameters, we begin with the mutual information between the predicted labels, $y$, and the categorical distribution, $\mu$, given as:
\begin{eqnarray}
    \mathcal{I}[y,\mu | x^{*}, \mathcal{D}] &=& \mathbbH\big[\mathbbE_{p(\mu|x^{*},\mathcal{D})}[p(y|\mu)]\big] \nonumber \\
    && \qquad - \mathbbE_{p(\mu | \mathbf{x}^{*},\mathcal{D})}\big[\mathbbH [p(y|\mu)]\big]\ .
\end{eqnarray}
The mutual information between the predicted labels and the categorical distribution quantifies how well the categorical distribution informs the label --- or how closely the assumed prior matches the data distribution. This uncertainty metric can then be decomposed into the total, aleatoric, and epistemic components. We first summarize the statistical theory underlying each of these quantities before introducing two complementary metrics which may also be defined in this setting, the {\it model knowledge uncertainty}, calculable from a KL divergence, and the {\it differential entropy}, which can be practically informative with respect to OOD behavior. \\

\noindent \textbf{Total Uncertainty.}
The total uncertainty is given by the entropy of the predictive posterior after marginalizing over the distribution, $\mu$, and the model parameters, $\theta$:
\begin{eqnarray}
    \mathbbH \big[\ex_{p(\mu | \mathbf{x}^{*},\hat{\theta})}[p(y|\mu)]\big] &=& - \sum_{c=1}^{C} \frac{\alpha_{c}}{\alpha_{0}} \ln \frac{\alpha_{c}}{\alpha_{0}}\ .
\end{eqnarray}

\noindent\textbf{Aleatoric Data Uncertainty.}
Data uncertainty measures the class overlap between labels. We can express the overlap between labels as the expected entropy of the predictions, $y$, given the distribution, $\mu$. This provides information about where the Dirichlet distribution is peaked. Quantitatively, the aleatoric uncertainty is
\begin{eqnarray}
    \ex_{p(\mu | \mathbf{x}^{*},\hat{\theta})}\big[\mathbbH [p(y|\mu)]\big] && \\
    &&\hspace{-2cm} =\sum_{c=1}^{C}-\frac{\alpha_{c}}{\alpha_{0}}\big(\psi(\alpha_{c} + 1) - \psi(\alpha_{0} + 1) \big)\ , \nonumber
\end{eqnarray}
where the object $\psi(\cdot)$ represent the digamma function. \\

\noindent\textbf{Distributional Knowledge (Epistemic) Uncertainty.}
Distributional knowledge or epistemic uncertainty is a measure of the spread of the Dirichlet distribution based on the underlying categorical distribution, $\mu$, which can be expressed through the mutual information: how well does the categorical distribution, $\mu$, inform the predicted labels, $y$, given the inputs, $x^{*}$, and the predicted parameters, $\hat{\theta}$. This quantity is non-existent in the Bayesian model-averaging framework; as such, its definition in an EDL context represents another useful source of information from the DPN. Explicitly, the distributional knowledge uncertainty may be evaluated from the difference of the total and aleatoric uncertainties. {\it Viz}.
\begin{eqnarray}
    \mathcal{I}[y,\mu | \mathbf{x}^{*}, \hat{\theta} ] &=& - \sum_{c=1}^{C} \frac{\alpha_{c}}{\alpha_{0}} \ln \frac{\alpha_{c}}{\alpha_{0}}  \\
    &-& \sum_{c=1}^{C}-\frac{\alpha_{c}}{\alpha_{0}}\big(\psi(\alpha_{c} + 1) - \psi(\alpha_{0} + 1) \big)\ . \nonumber
\end{eqnarray}

\noindent\textbf{Model Knowledge Uncertainty (KL Divergence).}
Model knowledge uncertainty is a measure of the spread of the Dirichlet distribution expressed as the mutual information between the predicted labels, $y$, and the model parameters, $\theta$, given as $\mathcal{I}[y,\theta | x^{*}, \mathcal{D} ] $. Since we are implementing the point approximation as $p(\theta | \mathcal{D}) = \delta(\theta - \hat{\theta})$, and are marginalizing over the categorical distribution $\mu$, this quantity is not calculable in the DPN framework in terms of Dirichlet parameters; however we can express the spread of the Dirichlet distribution through a complementary quantity, the expected KL divergence between two independent draws of categorical distributions from the Dirichlet prior:
\begin{eqnarray}
    \ex_{p(\mu^{1,2}|\mathbf{x}^{*},\hat{\theta})}\left [D_{KL}\Big (p(y|\mu^{1}) \Big \| p(y|\mu^{2})\Big ) \right] &=& \frac{C-1}{\alpha_{0}}\ .\ \ 
\end{eqnarray}

\noindent\textbf{Differential Entropy.}
Differential entropy measures the entropy of a continuous random variable and has no constraints on its value like the entropy of a discrete random variable does. The differential entropy may be evaluated as
\begin{eqnarray}
    \mathbbH [p(\mu | \mathbf{x}^{*},\hat{\theta})] &=& -\ln \frac{\Gamma(\alpha_{0})}{ \prod_{c=1}^{C}\Gamma(\alpha_{c})} \\
    &-& \sum_{c=1}^{C} (\alpha_{c}-1)\big(\psi(\alpha_{c} ) - \psi(\alpha_{0} ) \big)\ . \nonumber
\end{eqnarray}
The differential entropy is maximized when the Dirichlet distribution is flat, corresponding to OOD samples, and it therefore provides a sensitive measure of this behavior.

%

\section{Results}
\label{sec:results}

In this section, we detail the results of our DPN analysis to quantify overlaps among the different AEWI scenarios of Sec.~\ref{sec:bsm}--\ref{sec:pheno} in the two-dimensional $\Delta \chi^2$ latent space. We train our DPN on shifts in the $\Delta \chi^{2}$ statistics corresponding to variations in the CKM matrix elements, $V^\prime_{ij}$. These shifts are the three AEWI scenarios as described in Sec.~\ref{sec:bsm}. Our goal is to establish a quantitative metric for assessing the degree of separation or overlap between these models while offering insights into model distinguishability and commonality with respect to empirical data. Our results highlight the sensitivity of this method to subtle variations in electroweak parameters and its potential for probing BSM signatures.

For training data, we construct shifts in the CKM matrix elements representing three possible AEWI scenarios. Effectively, we dimensionally reduce these anomalous scenarios by projecting them to a tractable two-dimensional latent space derived from the calculated $\Delta \chi^{2}$ of the AEWI scenarios for $Q^{2} > 10$ and $Q^{2} \le 10$ GeV$^{2}$ against the CDHSW $F_{2}^{W^\pm}(x, Q^{2})$ data set using MC replicas of the NNLO CT18 PDFs. By formulating this effective $\Delta \chi^2$ space, we frame an interpretable coordinate system onto which the spread in our AEWI scenarios may be cast. The OOD samples for training are generated as random points close to the true data distributions in this two-dimensional $\Delta \chi^2$ space. These points correspond to separable scenarios in the larger space represented by our AEWI parametrization which are considered unexplored and which the DPN does not encounter during training.

Fig.~\ref{fig:dpn_demo} shows five distinct and representative example predictions of the DPN trained on the three shifts as described above inside the dimensionally reduced latent space. To illustrate, we choose five unique coordinates, each indicated by a large, red ``$\mathbf{\times}$'' in the $\Delta \chi^2$ plane (and labeled accordingly); for each such point, we predict the DPN samples indicated by the associated 2-simplexes with red dots as well as the full contour of the predicted Dirichlet distribution. 

Alongside these queries of the DPN within the $\Delta \chi^2$ plane, we also plot the original test data sets associated with the AEWI-induced variations to indicate regions where similar data have informed the DPN training; this representation permits a more human-readable interpretation of the Dirichlet distributions. In particular, the clusters of points in orange, blue, and green correspond to calculations base on each of the three different AEWI-induced shifts of the CKM matrix elements we label as Shift 1, Shift 2, Shift 3, respectively. 

From our five selected DPN predictions, we observe the expected behavior as described in Sec.~\ref{sec:dpn-theory}. The trained DPN predicts a sharp, highly-peaked distribution for a specific AEWI model when evaluated in a well-defined data region as seen for Queries {\bf 1} and {\bf 4} (associated with Shifts 1 and 3, respectively), while it has a flatter, more uniform distribution when sampled in regions of the $\Delta \chi^2$ for which there are no data, as is the case with Query {\bf 2}. Query {\bf 2} represents OOD sampling, for which the prediction coordinates are associated with a regime of $\Delta \chi^2$ unlike anything in the training set of AEWI-based CKM shifts and NNLO PDFs. In this instance, one can see in the Dirichlet samples (red dots) that the categorical distributions are completely diffuse and nearly perfectly uniform.

Another characteristic behavior is to be found in
regions of greater model overlap ({\it i.e.}, lying between two or more distinct concentrations of $\Delta \chi^2$ points), as is the case for Queries {\bf 3} and {\bf 5}. In these scenarios, there are two distinguishing behaviors. First, the Dirichlet distribution becomes more diffuse and uncertain about its coordinate within the simplex; in addition, the distribution is also {\it shifted} to a region on the simplex which does not represent a single unique class or AEWI shift scenario. For example, in the case of Query {\bf 3}, the distribution is located between Shifts 1 and 2, which produce $\Delta \chi^2$ arrays adjacent to the queried coordinate. The behavior is qualitatively similar with Query {\bf 5}, for which the Dirichlet is peaked between Shifts 2 and 3 --- also in agreement with the outlay of the sampled coordinate in the $\Delta \chi^2$ plane.

\begin{figure*}[!ht]
    \centering
    \includegraphics[width=0.93\columnwidth]{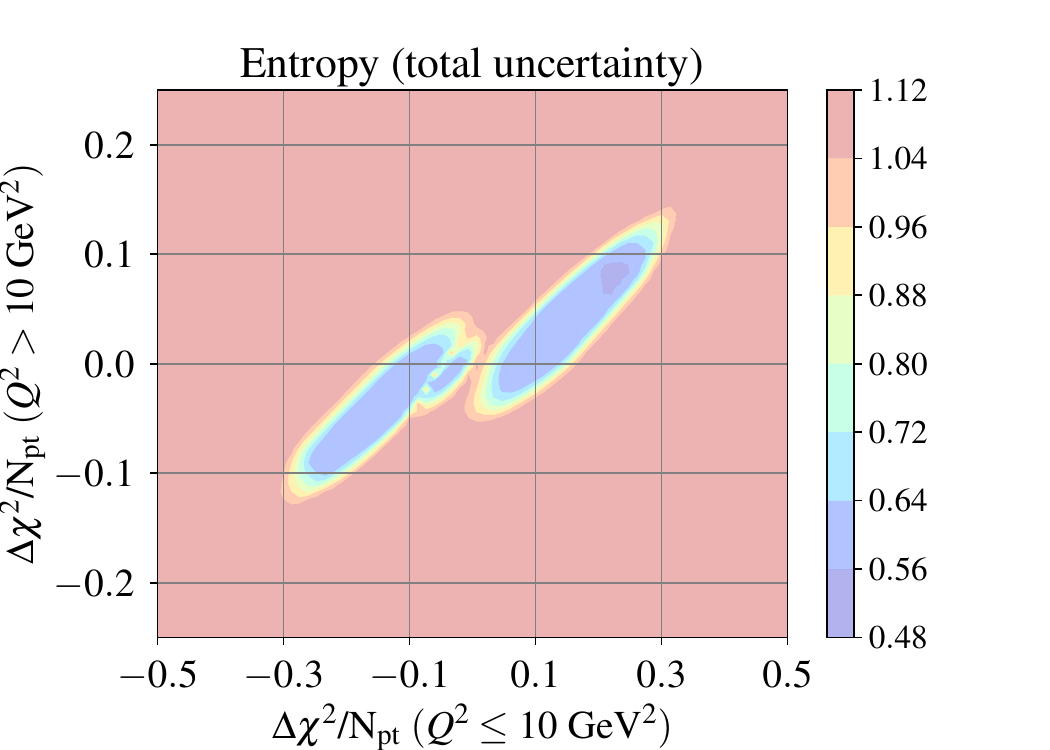}
    \includegraphics[width=0.93\columnwidth]{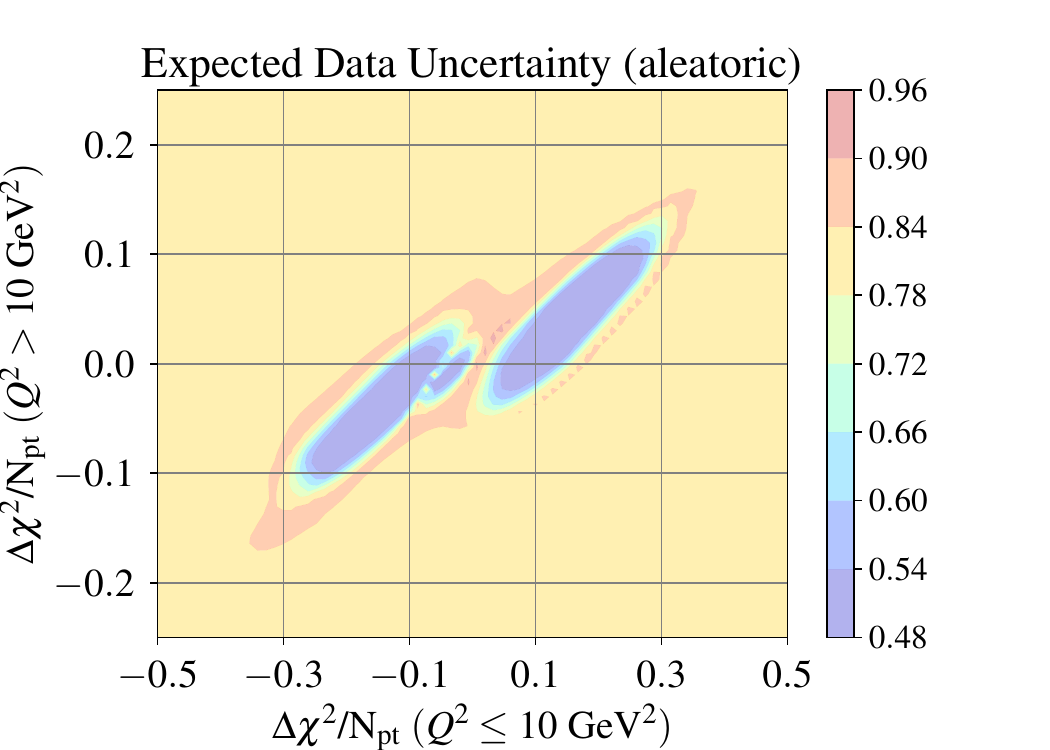} \\
    \includegraphics[width=0.93\columnwidth]{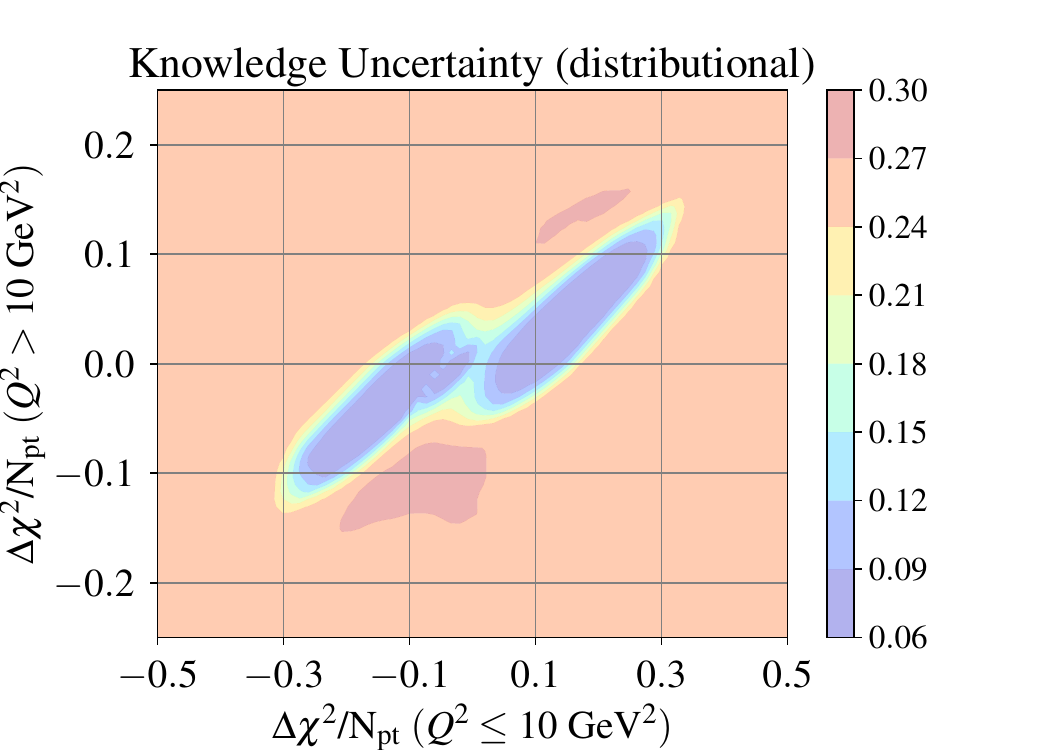}
    \includegraphics[width=0.93\columnwidth]{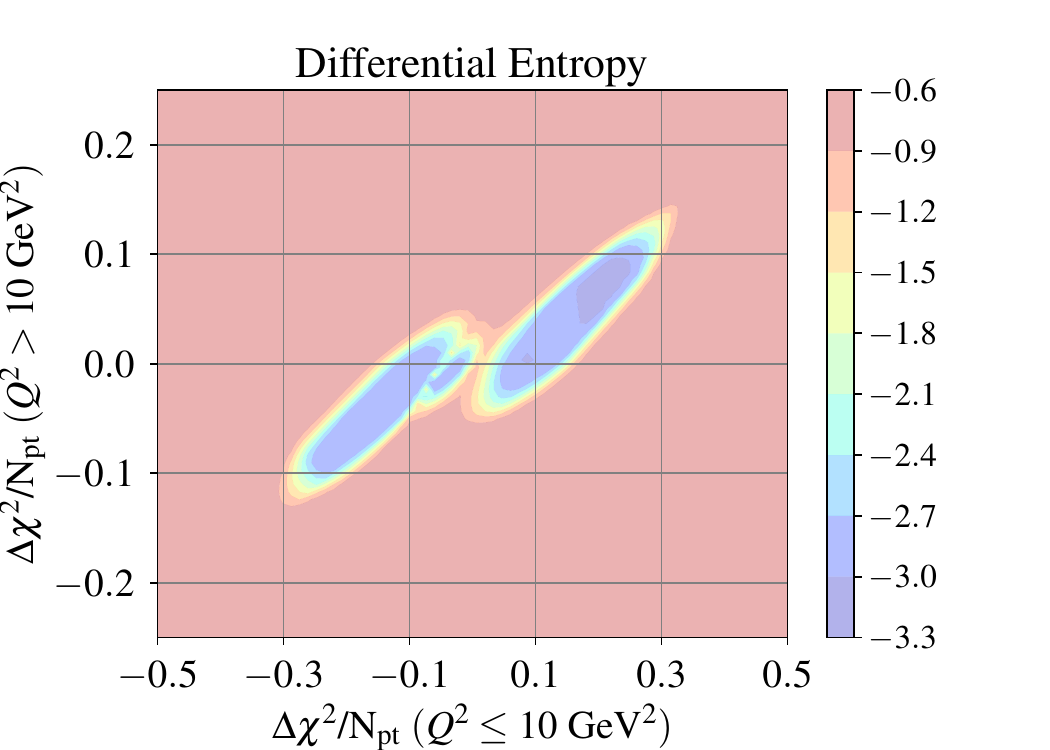}
    \includegraphics[width=0.93\columnwidth]{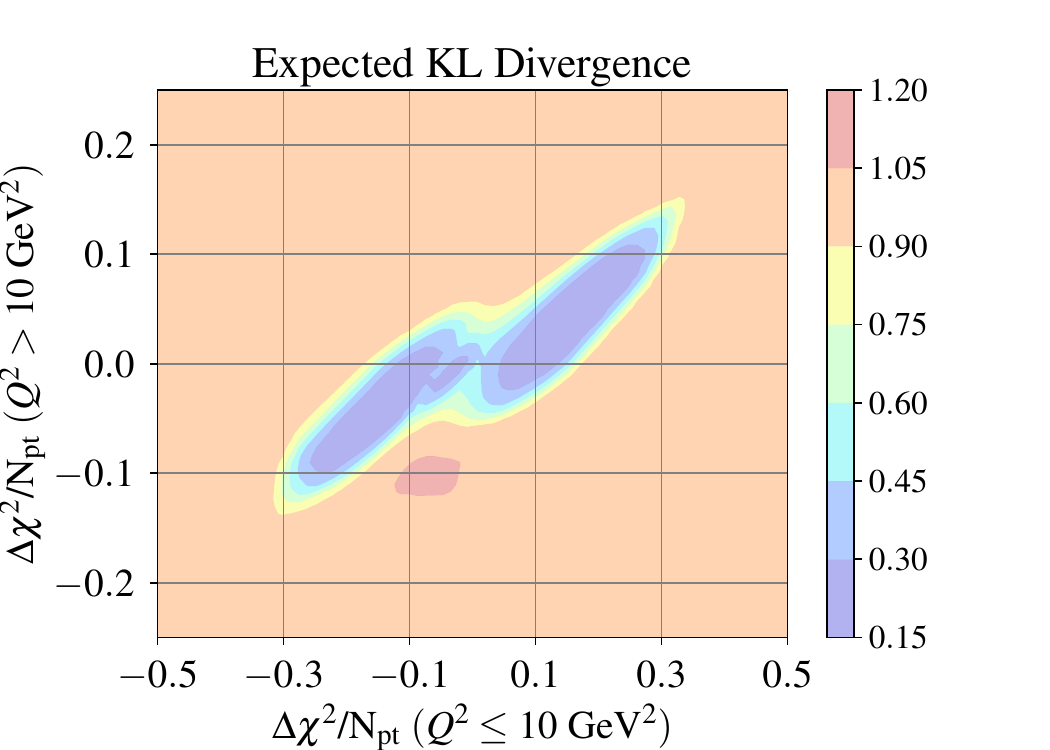}
    \caption{The measures of uncertainty as calculable in the DPN framework. Entropy (total uncertainty), expected data uncertainty (aleatoric), knowledge uncertainty (distributional), differential entropy, and expected KL divergence all are plotted as contours in a latent space calculated from the $\Delta \chi^{2} /\ $N$_\mathrm{pt}$  statistic for $Q^{2} > 10$ GeV$^{2}$ and $Q^{2} \le 10$ GeV$^{2}$ of the CDHSW $\nu$DIS data set for $F^\pm_{2}(x, Q^{2})$.}
    \label{fig:uq_ex}
\end{figure*}

In Fig.~\ref{fig:uq_ex}, we present complementary information by showing the uncertainty metrics of the classification task as detailed in Sec.~\ref{sec:uncertainty-metrics}. As noted in that section, there are five measures of uncertainty within the DPN framework: entropy (total) uncertainty; expected data (aleatoric) uncertainty; knowledge or epistemic (distributional) uncertainty; expected KL divergence; and differential entropy. These uncertainties may be calculated as functions of the coordinates in the two-dimensional $\Delta \chi^2$ latent space and depicted as contour maps. From these maps, it is possible to see general properties for each measure; in each case we identify through heat maps regions in which the uncertainty is lowest (shown in blue), which correspond to those sectors intersecting concentrated loci of training data. Meanwhile, regions where there are no data possess large uncertainties as indicated in red (with the one exception of the differential entropy, which has its own scale). By combining the information from all five measures of uncertainty, one can create an uncertainty profile that details exactly where each AEWI scenario lies in the $\Delta \chi^2$ plane with respect to the others.

\begin{itemize}
    \item The \textbf{total uncertainty} is a combination of the aleatoric and distributional (epistemic) uncertainties. It is approximately minimized by the aleatoric uncertainty. The total uncertainty gives a global picture of how well the DPN has modeled the underlying distribution of electroweak variations and classified each AEWI scenario --- statistically distinguishing it from the others. 
    \item \textbf{Aleatoric uncertainty} is given as the expected entropy of the predictive posterior. This is used as a metric to measure the class overlap of the AEWI scenarios. Notice that the aleatoric uncertainty is greatest at the boundaries of the training data; this aleatoric uncertainty generally increases both at the interfaces between AEWI-shift clusters in the training data, as well as at the boundaries between these data and OOD samples. The aleatoric uncertainty is nonzero in the region of the data because it is mapping the entropy of the training data distribution. 
    \item The \textbf{distributional (epistemic) uncertainty} is minimized in regions where the data lie, wherein it is approximately zero. This indicates that the DPN has accurately learned the probabilistic mapping from coordinate space to each specific AEWI scenario. It is not, however, a good indicator of OOD sampling, as the uncertainty value is much too low. The distributional uncertainty is a good indicator of where data exist which the model has already encountered during training.
    \item Next, we show the \textbf{expected KL divergence}, which is inversely proportional to the sum of the Dirichlet concentration parameters, $\alpha_{0}$. One can see that the distributional (epistemic) uncertainty and expected KL divergence are similar; this is because they are both approximately measuring the same quantity --- the spread in the categorical samples of the Dirichlet distribution.
    \item Lastly, to map the region for OOD sampling, we turn to the \textbf{differential entropy}. One can see that the differential entropy is allowed to be, and often is, negative by definition. There is a sharp distinction between data regions and OOD samples where the differential entropy is maximized (indicated in red). Due to how the OOD samples are defined and how evidence is collected in a DPN, there is not much structure to the differential entropy. Exploring more sophisticated methods built upon subjective logic is left for future work.
\end{itemize}

Another critical aspect of these uncertainty metrics is the boundary behavior as one transitions from a given region of the $\Delta \chi^2$ plane into another, which is related to extrapolation and interpolation. In Fig.~\ref{fig:uq_ex}, one can see that the  (distributional) knowledge or epistemic uncertainty dramatically increases outside the data region, although not immediately. There is a short extrapolation region, driven ultimately by the sparse ensemble of points along the periphery of the large density of data. This provides a buffer that gradually increases the knowledge uncertainty rather than a sharp cutoff. It should be expected that the knowledge uncertainty sharply increases outside the region of the training data, indicating that the model should be uncertain in this domain (even if it does predict a ``correct" result). Also, the data uncertainty in these boundary regions is much larger than in the OOD region. This is due to the collision of the data boundary and the OOD sampling boundary. This uncertainty between defining OOD and data induces a band of large data uncertainty surrounding the data samples before it reduces to a stable value.

By combining these five uncertainty metrics, it is possible to determine whether a given instance inside the $\Delta \chi^2$ plane is in-distribution with respect to a specific AEWI scenario, in-distribution but in a case of high overlap between AEWI models, or OOD with respect to the available AEWI-shifted theory scenarios. In-distribution within an AEWI scenario is indicated by low uncertainty across all five metrics; in-distribution but between scenarios produces high data- but low knowledge-uncertainty; and OOD queries possess high uncertainty across all five metrics. Using these measures, we have demonstrated that we can distinguish a representative collection of BSM-related AEWI scenarios in a common-setting latent space; we have further illustrated the UQ aspects given the ability to separate regions of large AEWI model overlap, indicating cases of parametric redundancy in the input theory. This theoretical technology permits the identification of regions in which high-impact BSM models might either be formulated or constrained in HEP searches for New Physics.

%

\section{Conclusions}
\label{sec:conclusions}

In this manuscript, we have for the first time introduced an EDL framework based upon DPNs for high-energy theory; we applied this method to quantify uncertainties involved in discriminating among variations in the electroweak sector; this approach allows us to map distinct configurations of anomalous electroweak interactions (AEWIs) as a basis for performing model discrimination.
Though general in scope, we concentrated our demonstration on the phenomenological problem of the BSM sensitivity of $\nu$DIS; operating on the presumption of a fixed PDF, we assumed several-$\sigma$ `discovery-level' AEWI-induced variations in the CKM matrix elements, and deployed our DPN methods to discriminate among these variations.
We note that these variations ultimately represent (sub)percent-level effects in the high-$x$ behavior of the $\nu$DIS structure functions (or reduced cross sections).
Our results demonstrate a successful separation of simplified AEWI scenarios as a proxy for a larger and richer space of specific BSM models; the results achieve this by measuring statistical overlaps among the various AEWI scenarios in an interpretable way through re-purposed EDL UQ techniques.

This result represents a step forward in the ML-based mapping of BSM models with UQ in an interpretable framework. While we focused on BSM scenarios generated through shifts in electroweak parameters --- specifically, the CKM matrix elements --- this approach may be extended arbitrarily to explore a wider range of BSM models or parametrizations of any parametric complexity, both within and beyond the neutrino sector.
In particular, the theoretical methods introduced in this manuscript are sufficiently general as to allow applications to global analyses involving BSM parametrizations (such as the EFTs discussed in Sec.~\ref{sec:bsm}), SM-only fits of PDFs, or simultaneous fits combining both elements. 
Regarding UQ, possible applications include improving interpretable latent representations through contrastive losses, incorporating generative methods for modeling BSM signatures, and finally building model similarity metrics into UQ and benchmarking studies to better understand parametrization dependence in fitting quantum correlation functions like the proton PDFs or related quantities.
By training DPN calculations like those shown above on samplings of theoretical predictions based on PDFs~\cite{Kovarik:2019xvh} or BSM parametrizations like EFTs~\cite{Costantini:2024wby}, the EDL methods of this study could also shed light on persistent questions in the treatment of uncertainties in each respective area.
Refining representations in embedding space by using contrastive losses will improve the interpretation of distances in latent spaces and allow for improved model separation and stability of similarity measures across training schemes.
In another class of possible extensions, generative methods like variational autoencoders (VAEs)~\cite{Kriesten:2023uoi}, can be used to both reduce the dimensionality of an input model space by optimizing the size of the latent space as a tunable parameter, and generate parametric models in regions of model overlap which are not in the training data set.
We envision a benchmarking scheme where this approach provides a quantitative way to measure congruence between models.
For example, the EDL methods in this study might be deployed on a more comprehensive set of specific BSM models --- in place of the AEWI scenarios taken as a proxy in this work; in this case, a union of generative methods like VAEs and the EDL techniques of this study could both interpolate among specific trained BSM models while also quantifying the uncertainties using metrics like those presented in Sec.~\ref{sec:uncertainty-metrics}.
Notably, tensions between models with significant overlap in observable space pose a significant challenge for uncertainty estimates and benchmarking.
In an effort to address the parametrization dependence of our results, it would be possible to apply the methods of this study to an expanded basis of electroweak or BSM model variations alongside simultaneously varying PDFs representing a wider range of assumptions.
We expect the EDL tools of this work to play a useful role in precisely such studies in future work. \\

%

\begin{acknowledgements}

TJH thanks members of CTEQ for helpful discussions on the subtleties of uncertainty definitions for global analyses related to the calculations shown in this study.
This work at Argonne National Laboratory was supported by the U.S.~Department of Energy under contract DE-AC02-06CH11357.

\end{acknowledgements}

\appendix
\label{app}

\section{AEWI-shifted CKM elements}

In this appendix, we explicitly list the entries of the shifted CKM matrix elements; as discussed in Sec.~\ref{sec:bsm}, we absorb the effects of anomalous electroweak interactions (AEWIs) directly into the CKM matrix elements as a proxy for BSM signatures as might be parametrized by EFTs or specific UV-complete models. In this work, we implement randomized deviations from PDG-preferred values~\cite{ParticleDataGroup:2024cfk} in three distinct scenarios. The CKM entries for these scenarios correspond to the elements of Eq.~(\ref{eq:Vprime}). In particular, we have
\begin{equation}
V^\prime_{ij}\big|_\mathrm{Shift\, 1} = \begin{pmatrix}
0.974159 & 0.225832 & 0.003782 \\
0.217707 & 0.975148 & 0.041112 \\
0.008533 & 0.041313 & 0.999110
\end{pmatrix}\ ;
\end{equation}
\begin{equation}
V^\prime_{ij}\big|_\mathrm{Shift\, 2} = \begin{pmatrix}
0.975272 & 0.220975 & 0.003827 \\
0.221022 & 0.974436 & 0.040291 \\
0.008756 & 0.040644 & 0.999135
\end{pmatrix}\ ;
\end{equation}
\begin{equation}
V^\prime_{ij}\big|_\mathrm{Shift\, 3} = \begin{pmatrix}
0.973485 & 0.228719 & 0.003844 \\
0.205910 & 0.977702 & 0.041239 \\
0.007999 & 0.040718 & 0.999139
\end{pmatrix}\ .
\end{equation}
%

%

\bibliographystyle{utphys}
\bibliography{cdhsw}

\end{document}